\documentclass[conference]{IEEEtran}
\IEEEoverridecommandlockouts
\usepackage{cite}
\usepackage{amsmath,amssymb,amsfonts}
\usepackage{graphicx}
\usepackage{textcomp}
\usepackage{xcolor}
\usepackage{float}
\usepackage{url}
\usepackage{tabularx}
\usepackage{slashbox}
\usepackage{multirow}
\usepackage{algorithm}
\usepackage{algorithmicx}
\usepackage{algpseudocode}
\usepackage{soul}
\usepackage{bbm}
\usepackage{setspace}
\usepackage[left=1.62cm,right=1.62cm,top=1.9cm,bottom=2.60cm]{geometry}
\newcommand\StateX{\Statex\hspace{\algorithmicindent}}

\DeclareMathOperator*{\argmin}{argmin}
\def\BibTeX{{\rm B\kern-.05em{\sc i\kern-.025em b}\kern-.08em
    T\kern-.1667em\lower.7ex\hbox{E}\kern-.125emX}}

\begin{document}

\title{Multi-Scenario Bimetric-balanced IoT Resource Allocation: An Evolutionary Approach\\
\thanks{$^{*}$Corresponding author: Yang Wang, yang.wang1@siat.ac.cn}}

\iffalse
\author{\IEEEauthorblockN{1\textsuperscript{st} Jiashu Wu}
\IEEEauthorblockA{\textit{SIAT \& University of CAS} \\
\textit{Chinese Academy of Sciences}\\
Shenzhen \& Beijing, China \\
js.wu@siat.ac.cn}
\and
\IEEEauthorblockN{2\textsuperscript{nd} Hao Dai}
\IEEEauthorblockA{\textit{SIAT \& University of CAS} \\
\textit{Chinese Academy of Sciences}\\
Shenzhen \& Beijing, China \\
hao.dai@siat.ac.cn}
\and
\IEEEauthorblockN{3\textsuperscript{rd} Yang Wang$^*$}
\IEEEauthorblockA{\textit{Shenzhen Inst of Adv Technology} \\
\textit{Chinese Academy of Sciences}\\
Shenzhen, China \\
yang.wang1@siat.ac.cn}
\and
\IEEEauthorblockN{4\textsuperscript{th} Zhiying Tu}
\IEEEauthorblockA{\textit{Harbin Institute of Technology} \\
Weihai, China \\
tzy\_hit@hit.edu.cn}
}
\fi

\author{\IEEEauthorblockN{Jiashu Wu$^{1,2}$, Hao Dai$^{1,2}$, Yang Wang$^{1,*}$, Zhiying Tu$^{3}$}
\IEEEauthorblockA{$^1$Shenzhen Institute of Advanced Technology, Chinese Academy of Sciences, Shenzhen 518055, China}
\IEEEauthorblockA{$^2$University of Chinese Academy of Sciences, Beijing 100049, China}
\IEEEauthorblockA{$^3$Harbin Institute of Technology, Weihai 264209, China}
}

\iffalse
\author{\IEEEauthorblockN{1\textsuperscript{st} Anonymous Authors}}
\fi

\maketitle

\begin{abstract}
In this paper, we allocate IoT devices as resources for smart services with time-constrained resource requirements. The allocation method named as \emph{BRAD} can work under multiple resource scenarios with diverse resource richnesses, availabilities and costs, such as the intelligent healthcare system deployed by Harbin Institute of Technology (HIT-IHC). The allocation aims for bimetric-balancing under the multi-scenario case, i.e., the profit and cost associated with service satisfaction are jointly optimised and balanced wisely. Besides, we abstract IoT devices as digital objects (DO) to make them easier to interact with during resource allocation. Considering that the problem is NP-Hard and the optimisation objective is not differentiable, we utilise Grey Wolf Optimisation (GWO) algorithm as the model optimiser. Specifically, we tackle the deficiencies of GWO and significantly improve its performance by introducing three new mechanisms to form the \emph{BRAD-GWA} algorithm. Comprehensive experiments are conducted on realistic HIT-IHC IoT testbeds and several algorithms are compared, including the allocation method originally used by HIT-IHC system to verify the effectiveness of the \emph{BRAD-GWA}. The \emph{BRAD-GWA} achieves a $3.14$ times and $29.6\%$ objective reduction compared with the HIT-IHC and the original GWO algorithm, respectively. 
\end{abstract}


%%%%%%%%%%%%%%%%%%%%
%%% Introduction %%%
%%%%%%%%%%%%%%%%%%%%

\section{Introduction}\label{sec:section_introduction}

With the rapid prevalence of ubiquitous IoT devices such as smart sensors and smart monitoring devices \cite{chettri2019comprehensive,10.1145/3394171.3413995,wu_iot}, applications such as healthcare \cite{iot_intelligent_healthcare_gandhi2018intelligent} and logistics \cite{iot_intelligent_logistics_wang2020intelligent} have became more intelligent. A real-life example is the Intelligent HealthCare system deployed by Harbin Institute of Technology (HIT-IHC), which involves various kinds of IoT devices such as smart gesture detector, smart treadmill, etc., to provide elders with healthcare services. The IoT smart space supports tremendous amount of ubiquitous services, which require different types of IoT devices as resources to accomplish in an effective way. However, as the number of services surges \cite{pecco,theta_iot}, allocating these resources in traditional way becomes labour-intensive and infeasible \cite{sangaiah2020iot,9833301} for IoT resource providers. 

The resource allocation for smart IoT services has gained gradual attentions. Several past research efforts \cite{zhao2019optimal,sangaiah2020iot} have been introduced to improve the resource allocation process with the goal of waiting time saving, budget minimisation, etc. However, these research efforts mainly focused on a single resource scenario, without considering the more realistic multi-scenario case, i.e., several resource scenarios with heterogeneous resource richnesses and costs. For example, the HIT-IHC system is deployed in the city hospital that represents resource-rich scenario, and community clinics that act as typical resource-scarce scenarios. Consequently, the bimetric-balanced resource allocation methods aiming to balance the service profit and cost under the multi-scenario case remained under-investigated. Besides, some research efforts utilised metaheuristic algorithms as their optimisers, however, as the metaheuristic algorithm quickly evolving, it is worth leveraging latest metaheuristic optimisers to make wiser allocation decisions. Finally, there was a void for past researches to verify the resource allocator on realistic IoT applications with multiple scenarios. 

To fill the voids of past research efforts, in this paper we consider a more sophisticated model named as \emph{BRAD}, which allocates IoT devices as resources in a \ul{B}imetric-balanced manner, i.e., balancing profit and cost wisely during \ul{R}esource \ul{A}llocation. Hence, it avoids the drawbacks of uni-metric methods, i.e., boosting profit while ignoring the cost, or reducing the cost while impairing the profit. Moreover, it can make bimetric-balanced decisions under multi-scenario case. Besides, the \emph{BRAD} utilises the concept of \ul{D}igital Object to abstract IoT devices to ease the interactions with IoT devices. 

Since the \emph{BRAD} optimisation model is computationally hard in nature \cite{hard1_tsai2018seira,sangaiah2020iot}, and its optimisation objective is not differentiable which prevents gradient-based optimisation methods from being used, we utilise the GWO algorithm \cite{grey_wolf_optimiser_mirjalili2014grey} to optimise the proposed \emph{BRAD} model. Compared with other metaheuristic algorithms, the GWO enjoys more effective optimisation performance and relatively faster convergence \cite{grey_wolf_optimiser_mirjalili2014grey}, hence is promising to make optimised allocation decisions. To further enhance the efficacy, we tackle some of its deficiencies such as dictated grey wolf social hierarchy, lacking grey wolf elimination mechanism, etc., and enhance its effectiveness by proposing an advanced version with three improvements: the bimetric-balanced and density-aware grey wolf initialiser, the enhanced greedy grey wolf social hierarchy mechanism and the lifetime-enabled grey wolf elimination mechanism. We name the proposed algorithm as \emph{BRAD-GWA} (A stands for ``advanced''). 

Finally, this paper evaluates the effectiveness of the \emph{BRAD-GWA} algorithm on the realistic HIT-IHC system and considers several comparing algorithms, including the allocation method originally used by the HIT-IHC system and the original GWO algorithm. \emph{BRAD-GWA} significantly outperforms its counterparts by a large margin, achieving $3.14$ times and $29.6\%$ objective reduction compared with the original HIT-IHC and the GWO algorithm, respectively. In summary, this paper makes contributions as follows: 

\begin{itemize}
  \item We propose a bimetric-balanced IoT resource allocation model, which can satisfy time-constrained resource needs under multi-scenario case. 
  \item We improve the Grey Wolf Optimisation algorithm using three mechanisms and substantially enhance its performance when optimising the \emph{BRAD} model. 
  \item We empirically apply the proposed \emph{BRAD-GWA} algorithm into a real intelligent healthcare system, i.e., HIT-IHC, and achieves significantly improved performance compared with its original resource allocation method. 
\end{itemize}

%The rest of the paper is organised as follows: Section \ref{sec:section_related_work} introduces related works regarding resource allocation efforts in smart space, as well as optimisation methods that can be used to address the problem, followed by our research opportunities. Backgrounds of our HIT-IHC intelligent healthcare system and the Grey Wolf Optimisation algorithm are provided in Section \ref{sec:section_background}. Section \ref{sec:section_model_and_method} presents the formulation of the \emph{BRAD} model. How the GWO is improved and integrated to form the advanced \emph{BRAD-GWA} is also presented in Section \ref{sec:section_model_and_method}. Section \ref{sec:section_implementation_discussion} discusses the implementation using the digital object abstraction. Section \ref{sec:section_experiment} provides the experimental setups, and analyses the experimental results to testify the effectiveness of the proposed algorithm. The last section concludes the paper. 

%%%%%%%%%%%%%%%%%%%%
%%% Related Work %%%
%%%%%%%%%%%%%%%%%%%%

\section{Related Work}
\label{sec:section_related_work}

\subsection{Resource Allocation for Services in Smart IoT Space}
\label{sec:section_resource_allocation_for_services_in_smart_space}

There are several research efforts focusing on metaheuristic-based IoT resource allocation methods. Jafari et al., \cite{jafari2021joint} utilised the Non-dominant Sorting Genetic Algorithm (NSGA) and the Bees Algorithm (BA) to schedule IoT resources for time-constrained tasks. They applied the overall energy consumption and time delay as the optimisation objective. Abdel-Basset et al., \cite{abdel2020energy} presented a QoS-oriented IoT resource-task scheduler based on Harris Hawks algorithm with a local search strategy. Jain et al., \cite{jain2021metaheuristic} tackled resource allocation for IoT environment via a Quasi Oppositional Search and Rescue Optimiser and achieved optimised overall system cost. Sangaiah et al., \cite{sangaiah2020iot} tackled the resource allocation for ever-growing amount of IoT services via the Whale Optimisation algorithm (WOA) with the total communication budget as the objective. Tsai et al., \cite{hard1_tsai2018seira} presented SEIRA algorithm, which was an extended version of Search Economics (SE) that can solve the IoT resource allocation problem with optimised communication cost. Several components were enhanced to improve the effectiveness of the algorithm, such as the solution encoding and the dynamic local search operator. 

However, these past research efforts still faced some deficiencies which need to be addressed. Firstly, these works only considered resource allocation in a single scenario, which degraded their effectiveness when working under the multiple heterogeneous resource scenarios. Besides, these works only considered costs, and ignored the profit brought by service satisfaction. Lacking bimetric-balancing may cause a profit-impaired and cost-oriented allocation strategy. Finally, these methods didn't consider the Grey Wolf optimiser, which experimentally outperforms some of the metaheuristic algorithms used in these research efforts. 

\subsection{Model Optimiser}
\label{sec:section_model_optimiser}

Numerous research efforts have been presented improve the efficacy of optimisers. Resende et al., \cite{grasp_resende2014grasp} presented the Greedy Randomised Adaptive Search Procedures (GRASP), which proposed the greedy randomised adaptive phase and the local search phase. However, it did not emphasise the balance between exploration and exploitation, the greedy approach may suffer from a higher chance to be trapped by the local optima. On the other hand, motivated by the natural phenomena, various metaheuristic algorithms were proposed. For instance, the Firefly Algorithm (FFA) \cite{firefly_algorithm_yang2010firefly} utilised the flashing patterns and attraction behaviours of fireflies to guide the firefly movement in the search space. The Whale Optimisation Algorithm (WOA) \cite{whale_optimisation_algorithm_mirjalili2016whale} was inspired by the bubble-net hunting strategy of whales and mimicked the social behaviours of humpback whales during hunting. The Particle Swarm Optimisation (PSO) algorithm \cite{kennedy1995particle_pso} was inspired by the social metaphor of artificial life in general, and conducted the computation in an evolutionised manner. There were some other algorithms got inspiration from physical phenomena in nature, such as the Gravitational Search Algorithm \cite{gravitational_search_algorithm_rashedi2009gsa}, Chemical Reaction Algorithm \cite{chem_algorithm_lam2009chemical}, etc. 

Despite the diversity of optimisation algorithms, according to the experimental results as in \cite{grey_wolf_optimiser_mirjalili2014grey}, these algorithms possessed inferior performance compared with the GWO algorithm. Empirically, the GWO can not only lead to solution that is more optimal, but also can converge more efficiently.

%%%%%%%%%%%%%%%%%%
%%% Background %%%
%%%%%%%%%%%%%%%%%%

\section{Background}
\label{sec:section_background}

In this section, we introduce the background of the HIT-IHC intelligence healthcare system to present the motivation of the problem. Then, the GWO algorithm will be introduced, its suitability and room of improvements are explained. 

\subsection{The HIT-IHC Intelligent HealthCare System}
\label{sec:section_hit_ihc_intelligent_healthcare_system}

\begin{figure}[!ht]
  \begin{center}
    \includegraphics[width=0.48\textwidth]{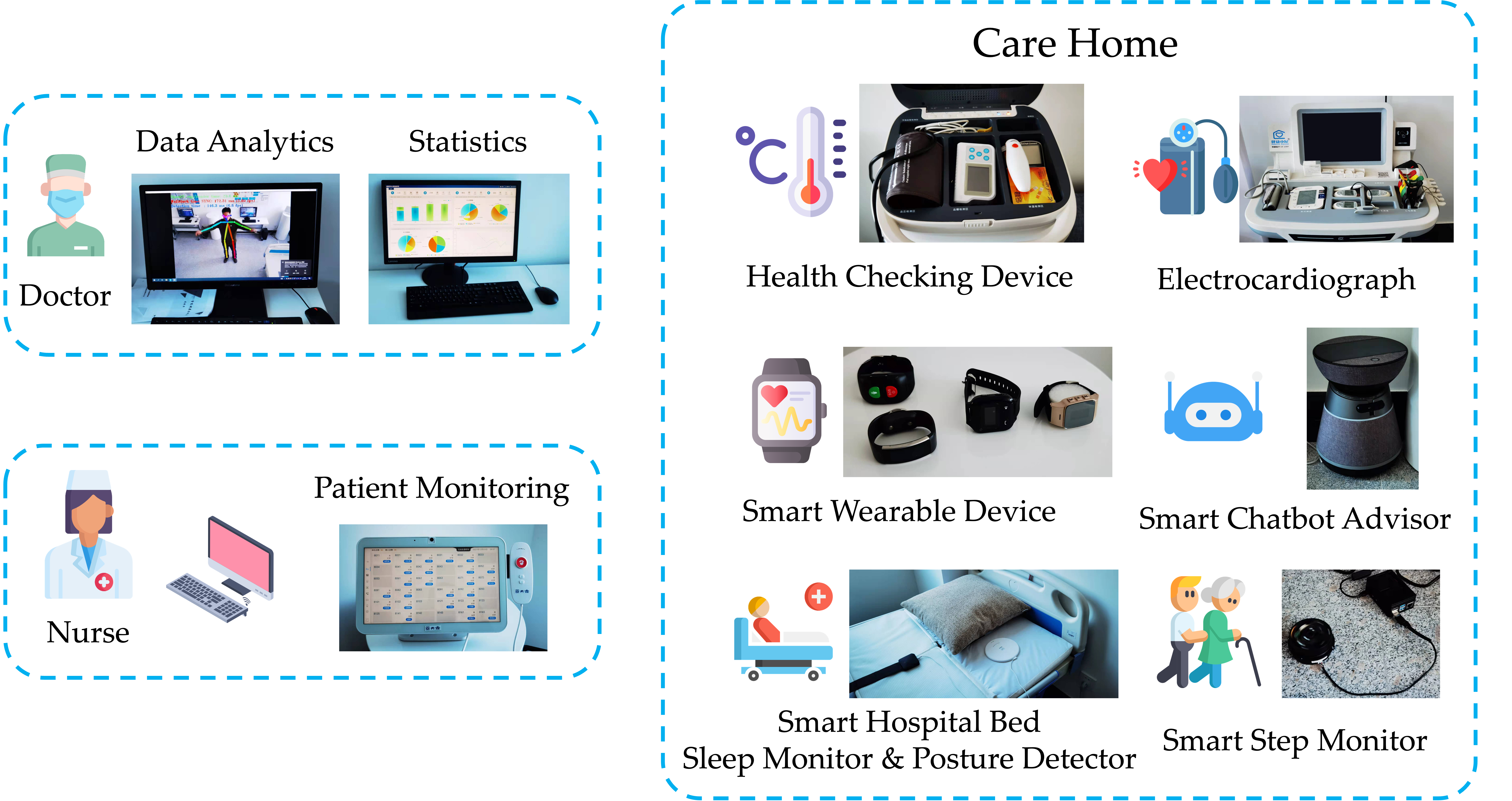}\\
    \caption{An illustration of the HIT-IHC system. Images are from a portion of real IoT devices of HIT-IHC system we deployed. }
    \label{fig:figure_intelligence_healthcare}
  \end{center}
\end{figure}
\vspace{-3mm}

The Harbin Institute of Technology Intelligent HealthCare system (HIT-IHC) is a system that utilises intelligent IoT devices to facilitate better elderly healthcare services. An illustration that contains real images of IoT devices of the HIT-IHC system have been shown in Figure \ref{fig:figure_intelligence_healthcare}. The HIT-IHC system contains around $30$ kinds of IoT devices that can assist the rehabilitation and treatment, each kind of IoT devices has $6$ to $100$ copies, with heterogeneous utility costs (device cost, energy consumption, etc.) and diverse allocated profit. The HIT-IHC system is deployed in the city hospital and community care home in the HIT campus in Shandong, China. Elders can request IoT devices such as health checking device, smart wearable device, etc., to assist the healthcare process. 

\textbf{Motivation} The HIT-IHC system currently applies a prioritised first come first serve method to allocate IoT devices to elderly with different needs, where HIT retired staffs are prioritised over other general users. The method favours the resource-rich scenario, i.e., the city hospital. If due to unavailability the request cannot be satisfied, the method will attempt to allocate it to the resource-scarce scenario, i.e., the local care home. However, under the multi-scenario environment, which possesses heterogeneous cost and profit, the method produces a very sparse resource allocation strategy with lots of unallocated short time slots and causes a lower resource utilisation efficiency. Besides, the method does not attempt to jointly optimise the cost and profit during resource allocation. Therefore, to improve the resource efficiency of the HIT-IHC system, a bimetric IoT resource allocation method that works under multiple heterogeneous scenarios is crucial, which motivates the proposal of the \emph{BRAD} model and the associated optimisation algorithm.

\subsection{The Grey Wolf Optimisation Algorithm}
\label{sec:section_the_grey_wolf_optimisation_algorithm}

The GWO algorithm is inspired by the organised social hierarchy of the grey wolf population\cite{grey_wolf_optimiser_mirjalili2014grey}. The ``alpha wolf ($X_\alpha$)'' is the leader, followed by the ``beta wolves ($X_\beta$)'' that assist alpha wolf. The ``omega wolves ($X_\omega$)'' reside at the bottom of the hierarchy and have to obey the command. Any wolves that are not alpha, beta or omega are classified as ``delta wolves ($X_\delta$)'', they follow the order of alpha and beta wolves, but can dominate omega wolves. Motivated by the hunting behaviours such as tracking, chasing and encircling the prey, the GWO algorithm treats the prey as the optimal value that is being searched for, and regards grey wolves as search candidates. By mimicking the hierarchical-based group hunting strategy, the GWO algorithm can cleverly approach the optima as much as possible. 

\textbf{Workflow} The GWO algorithm follows the workflow of swarm-based algorithms \cite{genetic_algorithms_framework_mirjalili2019genetic}. The species population will be initialised, and then keep evolving until reaching the termination condition, where the fittest individual will be regarded as the optima. 

\textbf{Formulation} The GWO algorithm utilises $K$ grey wolves acting as search candidates, each of them is a $d$ dimensional vector, where $d$ is the number of elements that the algorithm is optimising. Besides, each wolf is evaluated using a fitness function. The GWO algorithm randomly initialises the grey wolf population. Hence, the first deficiency is revealed. 

\ul{Deficiency 1: Random Initialisation without Density-awareness} Applying the random initialisation strategy will produce a population with low diversity in which two grey wolves can be highly similar. Consequently, the benefit of the population-based algorithm is hindered. 

In the GWO algorithm, grey wolves are search candidates, while the prey is not the real prey in real life. Instead, locations of dominated wolves act as the potential position of the prey. The alpha, beta and delta wolves can provide guidance during group hunting, while the bottom omega wolves provide no hunting knowledge at all. 

The GWO formulates the prey encircling mechanism as follows: 
\begin{equation}\label{equ:equation35}
  \begin{split}
    & D_{*} = \|C_{*} \times X_{*} - X\|, * \in \{\alpha, \beta, \delta\}
  \end{split}
\end{equation}
\begin{equation}\label{equ:equation36}
  \begin{split}
    & X_{*} = X_{*} - A_{*} \times D_{*}
  \end{split}
\end{equation}
\begin{equation}
  \begin{split}
    & A_{*} = 2 \alpha r_1 - \alpha, C_{*} = 2 r_2
  \end{split}
\end{equation}
\begin{equation}\label{equ:equation37}
  X(t + 1) = \frac{X_\alpha + X_\beta + X_\gamma}{3}
\end{equation}
where $t$ stands for the current iteration, components of $\alpha$ will linearly decrease from $2$ to $0$ over the course of iterations, and $r_1$, $r_2$ are random vectors in range $[0, 1]$. The rationales are as follows: given the setting of $\alpha$, $r_1$ and $r_2$, parameter vector $C$ will have range $[0, 2]$. $C > 1$ emphasises the effects of $\alpha$, $\beta$ and $\delta$ wolves in Equation \ref{equ:equation35}, letting these leading wolves to have a heavier influence of the next move of wolves and promotes exploitation. On the other hand, $C < 1$, de-emphasises the effect of leading wolves for better exploration. The degree of fluctuation of parameter vector $C$ remains unchanged during the course of iteration, hence, the GWO algorithm provides chances for exploration during the entire training process. As for parameter vector $A$, it has a fluctuation range between $[-2 \alpha, 2 \alpha]$, and the fluctuation range of $A$ keeps diminishing as $\alpha$ linearly decreases from $2$ to $0$ during the process. When $|A| < 1$, Equation \ref{equ:equation36} will let vector $X$ to be close to leading wolves and promotes exploitation. Conversely, if $|A| > 1$, vector $X$ will diverge from leading wolves for better exploration. By randomly generating parameter vector $A$ and $C$, the GWO algorithm can balance between exploration and exploitation. After each iteration, the best performed grey wolves will be elected as new $\alpha$, $\beta$ and $\delta$ wolves based on their fitness, and then they will provide guidance regarding probable position of the prey for other wolves to update their location in the next iteration based on Equation \ref{equ:equation35} to \ref{equ:equation37}. 

However, the design of the GWO algorithm still leaves room of improvements. 

\ul{Deficiency 2: Dictated social hierarchy} In GWO, the top-three performed grey wolves will be assigned as $\alpha$, $\beta$ and $\delta$ wolf, respectively. However, if some of these wolves are trapped by the local optima, then they will mislead other wolves to be trapped as well, leading to hindered optimisation performance. 

\ul{Deficiency 3: No lifetime elimination mechanism} In the design of the GWO algorithm, a grey wolf will keep updating no matter how bad its fitness is, while in nature, such kind of wolves is very likely to be eliminated by terminating their lifetime. Always letting these badly-performed grey wolves keep evolving will not yield positive influence towards the optimisation, especially during later rounds. 

\textbf{Advantage and Applicability} In summary, as a population-based algorithm, the GWO involves multiple search agents and balances between exploration and exploitation. Hence, it has higher chance to approximate the global optima. Besides, the GWO is gradient-free since it utilises a fitness function as the evaluator, which makes it especially applicable when the objective function is not differentiable. Finally, the GWO enjoys a faster convergence compared with its nature-inspired counterparts \cite{grey_wolf_optimiser_mirjalili2014grey}. 

\textbf{Improvement Opportunity} The GWO algorithm suffers from the aforementioned three deficiencies, it provides us with improvement opportunities. Therefore, we propose three new mechanisms, the bimetric-balanced and density-aware grey wolf initialiser that tackles deficiency 1, the enhanced greedy grey wolf social hierarchy that deals with deficiency 2, and lifetime-enabled grey wolf elimination mechanism that mitigates deficiency 3. Together, these mechanisms form the \emph{BRAD-GWA} algorithm. The details will be presented in Section \ref{sec:section_the_improvements}.

%%%%%%%%%%%%%%%%%%%%%%%%
%%% Model and Method %%%
%%%%%%%%%%%%%%%%%%%%%%%%

\section{Model and Method}
\label{sec:section_model_and_method}

\subsection{Problem Formulation}
\label{sec:section_problem_formulation}

In this section, we will firstly provide the problem formulation, then present the heterogeneous profit and cost model. 

\subsubsection{Problem and the \emph{BRAD} Model Formulation}
\label{sec:section_problem_and_the_brad_model_formulation}

Services in the IoT smart spaces can require diverse IoT resource to complete, and have their associated time constraint and profit gained. Utilising each resource will incur a cost. Due to different resource availabilities and costs in different scenarios, we consider the multi-scenario case, in which there is a resource-rich scenario ($\mathcal{R}$), as well as a resource-scarce scenario ($\mathcal{S}$) to reflect the real-world resource richness heterogeneities. 

In the \emph{BRAD} model, there are $M$ kinds of ubiquitous IoT resources, denoted as $R_m, m \in [1, M]$. Each resource is available in both the resource-rich and the resource-scarce scenario, denoted as follows:  
\begin{equation}
  R^X_m, X \in \{\mathcal{R}, \mathcal{S}\}, m \in [1, M]
\end{equation}
In this paper, we consider two scenarios, but the method can be extended to fit more scenarios in real world. Each resource $R^X_m$ can have $\alpha^X_m$ copies, the $j$\textsuperscript{th} copy of resource $R_m$ in scenario $X$ is denoted as follows: 
\begin{equation}
  R^X_{m}[j], X \in \{\mathcal{R}, \mathcal{S}\}, m \in [1, M], j \in [1, \alpha^X_m]
\end{equation}
Besides, each kind of resource, as well as each resource copy, can have a heterogeneous up and down time, denoted as $U(R^X_{m}[j])$ and $D(R^X_{m}[j])$, respectively. Meanwhile, we define the available period of resource $R^X_{m}[j]$ as $A(R^X_{m}[j])$, where $A$ stands for availability. Initially we have 
\begin{equation}
  A(R^X_{m}[j]) = [U(R^X_{m}[j]), D(R^X_{m}[j])]
\end{equation}
During the process, the available period $A(R^X_{m}[j])$ will be updated when a request of it is satisfied. 

In terms of the cost of IoT resources, each kind of resource $R^X_{m}$ can have heterogeneous cost per unit of time utilised, denoted as $C(R^X_{m})$. The same resource in resource-rich and resource-scarce scenarios can have different costs due to different resource richnesses and availabilities, while we assume that different copies of the same resource in the same scenario have the same cost. This cost model can model different utilisation costs caused by heterogeneities between scenarios. Moreover, it is also generalisable, as the cost it models can also jointly include various types of other costs, such as energy consumption, etc. 

As for services that require diverse resources to complete, there are in total $N$ of them. Each service $S_n$ can request $K$ kinds of resources to complete, where $1 \leq K \leq M$. We use $S^R_n$ to denote the set of resources required by service $S_n$, $S^i_n$ denotes the $i$\textsuperscript{th} resource required by service $S_n$, $S^{\#}_n$ denotes the number of resources service $S_n$ requires, and we have 
\begin{equation}\label{equ:resource_requirements_service_sn}
  \begin{split}
    & S^R_n = \{S^i_n, 1 \leq i \leq S^{\#}_n\}, S^i_n \in [1, M]
  \end{split}
\end{equation}
Each service is resource-time-constrained, therefore, we denote the starting and finishing time of each resource utilisation requested by service $S^i$ as $S(S^i_n)$ and $F(S^i_n)$, respectively. 

We assume the IoT service satisfaction to possess atomicity, i.e., either all its required resources are fully allocated based on the specified time constraints, or the entire service will be dissatisfied. This atomic service satisfaction mechanism prevents IoT resource deadlocks from happening. Moreover, we also assume the scenario-wise service satisfaction, i.e., the service will be entirely satisfied by resources in a single scenario. We define service resource request satisfaction as follows: 

\textbf{Definition 1 (IoT Resource request satisfaction)}: A IoT resource request $S^i_n$ of service $S_n$ is satisfied if \label{def:definition1}
\begin{equation}
  \begin{split}
    & \exists R^X_{S^i_n}[j] \in R^X_{S^i_n}, X \in \{\mathcal{R}, \mathcal{S}\}, j \in [1, \alpha^X_{S^i_n}],\\
    & s.t. [S(S^i_n), F(S^i_n)] \in A(R^X_{S^i_n}[j])
  \end{split}
  \label{equ:equation_single_resource_request_satisfied}
\end{equation}

Equivalently, we define 
\begin{equation}\label{equ:equation_single_resource_request_satisfied_indicator_function}
  1_{(S^i_n, X, j)}, X \in \{\mathcal{R}, \mathcal{S}\}, j \in [1, \alpha^X_{S^i_n}]
\end{equation}
which will return $1$ if the resource request $S^i_n$ can be satisfied by the $j$\textsuperscript{th} copy of resource $R_{S^i_n}$ in scenario $X$, i.e., Equation \ref{equ:equation_single_resource_request_satisfied} in Definition $1$ holds. Therefore, we define the service satisfaction as follows: 

\textbf{Definition 2 (Service satisfaction)}: The service $S_n$ is satisfied if the following holds: \label{def:definition2}
\begin{equation}\label{equ:equation_service_all_satisfied}
  \begin{split}
    & \forall S^i_n \in S^R_n, \exists j_{S^i_n} \in [1, \alpha^X_{S^i_n}], X \in \{\mathcal{R}, \mathcal{S}\},\\
    & s.t. 1_{(S^i_n, X, j_{S^i_n})} = 1
  \end{split}
\end{equation}
which is equivalent as for all resource request $S^i_n$ in $S^R_n$ of service $S_n$, there exist a $(X, j)$ combination that makes Equation \ref{equ:equation_single_resource_request_satisfied_indicator_function} in Definition $2$ hold. 
Hence, we then define 
\begin{equation}\label{equ:equation_service_resource_satisfied}
  1_{(S^R_n, X)}, X \in \{\mathcal{R}, \mathcal{S}\}
\end{equation}
which will return $1$ if Equation \ref{equ:equation_service_all_satisfied} holds, where for all resource requests, $X$ should be the same scenario, it cannot be satisfied by more than one scenarios simultaneously. The Equation \ref{equ:equation_service_resource_satisfied} can then be simplified as follows: 
\begin{equation}
  1_{(S_n, X)}, X \in \{\mathcal{R}, \mathcal{S}\}
\end{equation}
if this service can be satisfied in scenario $X$. 

To model the profit, we apply a heterogeneous profit model. Each service $S_n$ will produce different profit when being satisfied in different scenarios, which is denoted as 
\begin{equation}
  P(S_n, X), X \in \{\mathcal{R}, \mathcal{S}\}
\end{equation}

Finally, in terms of the multi-service IoT resource allocation strategy, we denote the allocation using $\mathcal{A}$. The initial allocation and the optimised allocation are denoted as $\mathcal{A}^I$ and $\mathcal{A}^O$, respectively. The allocation $\mathcal{A}$ is defined as follows: 
\begin{equation}
  \mathcal{A} = \begin{bmatrix}
    \mathcal{A}_{S_1}\\
    \mathcal{A}_{S_2}\\
    \vdots\\
    \mathcal{A}_{S_N}
  \end{bmatrix} = \begin{bmatrix}
    \begin{pmatrix}
      \mathcal{A}_{S^1_1, j_1} & \mathcal{A}_{S^2_1, j_2} & \cdots & \mathcal{A}_{S^{K_1}_1, j_{K_1}}
    \end{pmatrix}\\
    \begin{pmatrix}
      \mathcal{A}_{S^1_2, j_1} & \mathcal{A}_{S^2_2, j_2} & \cdots & \mathcal{A}_{S^{K_2}_2, j_{K_2}}
    \end{pmatrix}\\
    \vdots\\
    \begin{pmatrix}
      \mathcal{A}_{S^1_N, j_1} & \mathcal{A}_{S^2_N, j_2} & \cdots & \mathcal{A}_{S^{K_N}_N, j_{K_N}}
    \end{pmatrix}
  \end{bmatrix}
\end{equation}
where $\mathcal{A}_{S^2_1, j_3}$ means the second resource required by service $S_1$ is satisfied by the $j_3$\textsuperscript{th} copy of that resource. 

\subsection{The \emph{BRAD} Profit and Cost Model}
\label{sec:section_the_brad_profit_and_cost_model}

\subsubsection{Heterogeneous \emph{BRAD} Profit Model}
\label{sec:section_heterogeneous_brad_profit_model}

In the \emph{BRAD} model, the heterogeneous profit model is defined as follows: 
\begin{equation}\label{equ:equation_profit}
  \begin{split}
    & P(S, R, \mathcal{A}) =\sum^N_n P(S_n, \mathcal{A}_{S_n})\\
    & = \sum^N_n (1_{(S_n, \mathcal{R})} \times P(S_n, \mathcal{R}) + 1_{(S_n, \mathcal{S})} \times P(S_n, \mathcal{S}))
  \end{split}
\end{equation}
where the indicator function $1_{S_n, X}$ will return $1$ if service $S_n$ is satisfied in scenario $X$. If service $S_n$ is not satisfied due to IoT resource shortage, then two indicator functions will all be $0$, i.e., no profit gained. 

The heterogeneous profit is generalisable since it can not only model the revenue produced by satisfying a service, but also can indirectly reflect other factors such as the costs of migrating a service from one scenario to the other, e.g., transferring a patient from the community clinic to the hospital due to resource unavailability. The cost of transferring can be implicitly modelled in the profit by deducing the profit correspondingly. Hence, the heterogeneous profit model is generalisable to include other factors. 

\subsubsection{Heterogeneous \emph{BRAD} Cost Model}
\label{sec:section_heterogeneous_brad_cost_model}

The \emph{BRAD} optimisation model also applies a heterogeneous resource cost model, which not only considers different utilisation costs caused by scenario-wise heterogeneities, but also can integrate other costs implicitly, such as energy consumption, computation costs, etc. Hence, the generalisability offered by the heterogeneous cost model eases the effort of considering more cost models explicitly. We define the utilisation period length function which returns the length of the utilisation period for service $S_n$ when requesting resource $S^i_n \in S^R_n$ as follows: 
\begin{equation}
  len: \mathbb{R} \rightarrow \mathbb{R}, len(S^i_n) = F(S^i_n) - S(S^i_n)
\end{equation}
and hence, the overall cost of all services is defined as follows: 
\begin{equation}\label{equ:equation_cost}
  \begin{split}
    & C(S, R, \mathcal{A}) = \sum^N_n C(S_n, R, \mathcal{A}_{S_n})\\
    & = \sum^N_n (1_{(S_n, \mathcal{R})} \times \sum_{S^i_n \in S^R_n} len(S^i_n) * C(R^\mathcal{R}_{S^i_n})\\
    & + 1_{(S_n, \mathcal{S})} \times \sum_{S^i_n \in S^R_n} len(S^i_n) * C(R^\mathcal{S}_{S^i_n}))
  \end{split}
\end{equation}

\subsubsection{Bimetric-balanced Optimisation Objective}
\label{sec:section_overall_optimisation_objective}

Finally, the \emph{BRAD} optimisation model will jointly optimise both profit and cost to form a bimetric-balanced model. To mask the range differences between profit and cost and balance them properly during optimisation, a negative scalar factor is multiplied on the profit. The bimetric-balanced optimisation objective is defined as follows: 
\begin{equation}\label{equ:overall_objective}
  Obj(S, R) = \argmin_{\mathcal{A}^O}\{C(S, R, \mathcal{A}) + w \times P(S, R, \mathcal{A})\}
\end{equation}

Optimising the objective function of the \emph{BRAD} model can jointly optimise both profit and cost as much as possible and hence provide bimetric-balanced allocation decision for IoT resource providers. 

\subsection{The \emph{BRAD-GWA} Algorithm}
\label{sec:section_the_grey_wolf_optimiser}

In this section, we introduce three improvements made to improve the GWO algorithm, and how the GWO algorithm is integrated to tackle the \emph{BRAD} model, e.g. what grey wolves stand for and how services are allocated based on the yielded allocation strategy $\mathcal{A}^O$, etc. 

\subsubsection{Algorithm Improvements}
\label{sec:section_the_improvements}

We propose an advanced Grey Wolf Optimiser called \emph{BRAD-GWA} which tackles the deficiencies as mentioned in Section \ref{sec:section_the_grey_wolf_optimisation_algorithm} and therefore boost the performance. 

\begin{algorithm}[!ht]
  \begin{algorithmic}[1]
      \Require
          \StateX Number of grey wolves $nsa$, 
          \StateX Allocation upper bound $ub$
      \Ensure grey wolf initialisation matrix with dimension $nsa \times N$
      \For{$S_n$ \textup{in} $S_N$}
          \State Calculate the profits, costs and bimetric objective of satisfying service $S_n$ in both scenarios using Eq. \ref{equ:equation_profit}, \ref{equ:equation_cost} and \ref{equ:overall_objective}
      \EndFor
      \For{$i$ \textup{in} $range(nsa \times 1.5)$}
          \For{$n$ \textup{in} $range(N)$}
              \If{$Obj(S_n, R)^\mathcal{R} \leq Obj(S_n, S)^\mathcal{S}$}
                  \State Store random value in range $[0, \frac{ub}{2})$ into $\mathcal{A}^I_i$
              \Else
                  \State Store random value in range $[\frac{ub}{2}, ub]$ into $\mathcal{A}^I_i$
              \EndIf
          \EndFor
      \EndFor
      \While{$len(\mathcal{A}^I) \neq nsa$}
          \State Find pair $(\mathcal{A}^I_i, \mathcal{A}^I_j)$ with minimum pairwise $L2$ distance
          \State Add $\frac{\mathcal{A}^I_i + \mathcal{A}^I_j}{2}$ into $\mathcal{A}^I$
          \State Remove both $\mathcal{A}^I_i$ and $\mathcal{A}^I_j$ from $\mathcal{A}^I$
      \EndWhile
      \State \Return $\mathcal{A^I}$
  \end{algorithmic}
\caption{The bimetric-balanced and density-aware grey wolf initialiser $initialiser(nsa, S_N, ub)$}
\label{algo:the_profit_cost_density_aware_initialisation}
\end{algorithm}

\textbf{Improvement 1 (Bimetric-balanced and Density-aware Grey Wolf Initialiser)}: We design a new grey wolf initialiser that is bimetric-balanced and density-aware to tackle the \ul{Deficiency 1: random initialisation without density-awareness}, as shown in Algorithm \ref{algo:the_profit_cost_density_aware_initialisation}. 

\ul{Bimetric-balanced}: The new initialiser allocates services (grey wolves) to the scenario with lower bimetric-balanced objective value. Allocation in the reversed way can result in service transfers, which causes extra costs. Hence the bimetric-balanced initialiser should outperform its random counterpart. Its mechanism has been shown in line $4$ - $12$ in Algorithm \ref{algo:the_profit_cost_density_aware_initialisation}. 

\ul{Density-awareness}: The random initialisation ignores the population density, i.e., it may randomly generate multiple individuals with high similarity and therefore hinders the effectiveness of population-based optimisers. The new initialiser generates more grey wolf vector than required, and iteratively merge the closest pair of grey wolf vectors by taking the average of them. The initialiser has been given in line $13$ - $17$ in Algorithm \ref{algo:the_profit_cost_density_aware_initialisation}. The merging will continue until the number of grey wolf vectors reaches the required number. By being density-aware, the population diversity is increased and hence improve the chance for the population-based optimiser to better approximate the global optima. 

\begin{algorithm}[!ht]
  \begin{algorithmic}[1]
      \Require Number of search candidate (grey wolves) $nsa$
      \State $\tau \leftarrow \frac{CI}{MI}$, $\tau$ denotes the elimination threshold
      \State Initialise $bl$ to be a dictionary which tracks the lifetime beginning of each grey wolf and is initialised to be $0$
      \For{$i$ \textup{in} $[0, nsa]$}
          \If{\textup{grey wolf} $G_i$ \textup{has the worst fitness among all grey wolves}}
              \State $e \leftarrow random(0, 1)$
              \If{$e \geq \tau$}
                  \State Set $bl[i] \leftarrow CI$
                  \If{$random(0, 1) \leq 0.5$}
                      \State $X^{[i]}(t + 1) \leftarrow$ a randomly initialised new grey wolf vector
                  \Else
                      \State $X^{[i]}(t + 1) \leftarrow lb + ub - G[random(0, nsa)]$
                  \EndIf
              \EndIf
          \Else
              \State $a[i] \leftarrow 2 - \frac{2(CI - bl[i])}{MI - bl[i]}$
              \State $\delta w[i] \leftarrow 1 - \frac{CI - bl[i]}{MI - bl[i]}$
              \For{$j$ \textup{in} $[1, 7]$}
                  \State Calculate $D_j \leftarrow \|C_j \times X_{\mathbbm{j}} - X\|$ (in Eq. (\ref{equ:equation35}))

                  \State Calculate $X_j \leftarrow X_j - A_j \times D_j$ (in Eq. (\ref{equ:equation36}))
          
                  \State Calculate $A \leftarrow 2r_1 \times a[i] - a[i]$
          
                  \State Calculate $C \leftarrow 2r_2$
          
                  \State $X^{[i]}(t+1)_A \leftarrow \frac{X_\alpha + \frac{\sum_{*=1}^2 X_{\beta*}}{2} + \frac{\sum_{*=1}^3 X_{\delta*} \times \delta w[i]}{3}}{2 + \delta w[i]}$
          
                  \State $X^{[i]}(t+1)_B \leftarrow \frac{X_{\alpha1} + \frac{\sum_{*=1}^2 X_{\beta*}}{2} + \frac{\sum_{*=1}^3 X_{\delta*} \times \delta w[i]}{3}}{2 + \delta w[i]}$
          
                  \State $X^{[i]}(t+1)_C \leftarrow \frac{X_{\alpha2} + \frac{\sum_{*=1}^2 X_{\beta*}}{2} + \frac{\sum_{*=1}^3 X_{\delta*} \times \delta w[i]}{3}}{2 + \delta w[i]}$
          
                  \State $X^{[i]}(t+1)_D \leftarrow $
                  \StateX $\frac{\frac{\sum_{*=1}^2 X_{\alpha*}}{2} + \frac{\sum_{*=1}^2 X_{\beta*}}{2} + \frac{\sum_{*=1}^3 X_{\delta*} \times \delta w[i]}{3}}{2 + \delta w[i]}$

                  \State $X^{[i]}(t+1) \leftarrow$ the one with best fitness among $(X^{[i]}(t+1)_A, X^{[i]}(t+1)_B, X^{[i]}(t+1)_C, X^{[i]}(t+1)_D)$
              \EndFor
          \EndIf
      \EndFor
  \end{algorithmic}
\caption{The enhanced greedy grey wolf social hierarchy and the lifetime-enabled grey wolf elimination mechanism}
\label{algo:the_improvement23_gwa}
\end{algorithm}

\textbf{Improvement 2 (Enhanced Greedy Grey Wolf Social Hierarchy)}: To tackle the \ul{Deficiency 2: Dictated Social Hierarchy}, an enhanced grey wolf social hierarchy with greedy strategy is applied. The original GWO utilised top-three best-performed grey wolves as $\alpha$, $\beta$ and $\delta$ wolves, which then influence how other wolves are updated. If they are trapped by local optima, then other wolves will be misled to be trapped as well, which significantly impairs the effectiveness of the algorithm. Given that utilising the top-three wolves is risky to suffer from local optima stagnation, in the enhanced grey wolf social hierarchy, top-seven best-performed grey wolves are utilised to avoid dictation from happening. The enhanced social hierarchy has been indicated in Algorithm \ref{algo:the_improvement23_gwa}. 

To further avoid dictation, i.e., a grey wolf being trapped by the local optima has a heavy influence on other wolves, four different social hierarchies are applied. For instance, in social hierarchy $A$ as indicated in line $22$ in Algorithm \ref{algo:the_improvement23_gwa}, the best grey wolf will be treated as $\alpha$ wolf, grey wolves ranked second and third are regarded as $\beta$ wolves ($\beta1-2$), and grey wolves ranked from $4$ to $6$ are regarded as $\delta$ wolves ($\delta1-3$). On the other hand, in social hierarchy $B$, $C$, and $D$, the top-two grey wolves are regarded as $\alpha$ wolves ($\alpha1-2$), grey wolves ranked from $3$ to $4$ are regarded as $\beta$ wolves ($\beta1-2$), and grey wolves ranked from $5$ to $7$ are treated as $\delta$ wolves ($\delta1-3$). In social hierarchy $B$, $C$ and $D$, either the first $\alpha$ wolf ($\alpha1$), or the second $\alpha$ wolf ($\alpha2$), or the combination of them will be utilised to avoid dictation as much as possible. 

Finally, a greedy strategy is leveraged to select the social hierarchy that yields the best fitness to produce guidance for other grey wolves to update their position. 

\ul{Exploration and Exploitation Balance}: Two levels of efforts have been put through to balance between exploration and exploitation in the enhanced greedy grey wolf social hierarchy, i.e., intra-hierarchy and inter-hierarchy. 

Each grey wolf has its own randomly initialised $A$ and $C$ parameter vectors, hence, different grey wolves may greedily use different social hierarchies, which introduces diversity of social hierarchies to promote better exploration and will lead to better optimisation outcome. Inside each social hierarchy, a delta weight $\delta w$ that linearly decrease from $1$ to $0$ during the course of iteration has been applied on $\delta$ wolves. During the initial iteration stage, $\delta$ wolves will be applied with a heavier weight to promote better exploration. As the iteration evolves, each grey wolf is likely to find its better prey position, hence, the influence of the $\delta$ wolves are gradually diminished to encourage each wolf to pursue its own favoured location, i.e., gradually emphasising exploitation. By balancing exploration and exploitation properly, the advanced \emph{BRAD-GWA} algorithm can better approximate the global optima. 
 
\textbf{Improvement 3 (Lifetime-enabled Grey Wolf Elimination Mechanism)}: To solve the \ul{Deficiency 3: lack of lifetime elimination mechanism}, we design a lifetime-enabled grey wolf elimination mechanism inspired by the natural elimination in nature. 

As indicated in line $5$ to $13$ in Algorithm \ref{algo:the_improvement23_gwa}, the grey wolf with the worst fitness will be considered for elimination. However, in nature, the worst-performed grey wolf is not necessarily eliminated immediately when it becomes the worst one. Hence, in line $2$ in Algorithm \ref{algo:the_improvement23_gwa}, a dynamic elimination threshold $\tau$ is applied, which linearly increases from $0$ to $1$ during the course of iterations. The rationale is that at the beginning of the training process, performances of all grey wolves are not stable enough. Hence, turn on the elimination too early may hinder the exploration effect. As the training evolves, especially during later period, if the grey wolf is still not promising, it becomes worthy to be eliminated. Therefore, as in line $6$ to $7$ in Algorithm \ref{algo:the_improvement23_gwa}, a decider $e$ will be a random value between $0$ and $1$, and the worst-performed grey wolf will only be eliminated if the decider is higher than the elimination threshold. Upon being eliminated, since parameter $a$ is a value keeps linearly decreasing from the beginning of the training process, the elimination will reset the decreasing process by updating the begin lifetime tracker $bl$ in line $8$ in Algorithm \ref{algo:the_improvement23_gwa}. Besides, another random value will be generated. If less than or equal to $0.5$, the eliminated grey wolf will be randomly initialised to start a new lifetime. Otherwise, the eliminated grey wolf will be set as the opposition of a randomly selected survived grey wolf to promote exploration to the location that is opposed to the selected grey wolf. The opposition-based lifetime renewal has been indicated in line $12$ in Algorithm \ref{algo:the_improvement23_gwa}. 

Hence, by utilising the lifetime-enabled grey wolf elimination mechanism, unpromising grey wolves will be gradually eliminated, especially during later training stage, which promotes exploration. 

\iffalse % removed due to space limitation
\begin{algorithm}[!ht]
  \begin{algorithmic}[1]
      \Require 
          \StateX Number of search candidate (grey wolf) $nsa$, 
          \StateX Allocation upper bound $ub$, 
          \StateX Objective function $Obj()$ as defined in Eq. (\ref{equ:overall_objective})
      \Ensure IoT Resource allocation strategy $\mathcal{A}^O$ for multi-services, which is the best grey wolf $G_\mathbbm{1}$
      \State Initialise the grey wolf population $G$ using Algorithm \ref{algo:the_profit_cost_density_aware_initialisation}
      \State Initialise parameter $a$, $A$ and $C$
      \State Calculate the fitness of each grey wolf using the objective function $Obj()$ (Eq. (\ref{equ:overall_objective}))
      \State Select the top-seven best-performed grey solves to act as upper hierarchy
      \While{$MI$ \textup{is not reached}}
          \State Utilise the enhanced greedy grey wolf social hierarchy as in Algorithm \ref{algo:the_improvement23_gwa} to update the position of each grey wolf
          \State Trigger elimination as indicated in Algorithm \ref{algo:the_improvement23_gwa}
          \State $CI \leftarrow CI + 1$
      \EndWhile
      \State \Return the best grey wolf
  \end{algorithmic}
\caption{Workflow of the \emph{BRAD-GWA} Algorithm}
\label{algo:the_brad_gwa_optimisation_algorithm}
\end{algorithm}
\fi

\textbf{Overall Workflow of the \emph{BRAD-GWA} Algorithm} 

%The overall workflow of the \emph{BRAD-GWA} algorithm has been shown in Algorithm \ref{algo:the_brad_gwa_optimisation_algorithm}. Firstly, the algorithm will utilise the proposed bimetric-balanced and density-aware grey wolf initialiser to initialise the grey wolf population. Meanwhile the algorithm will initialise the parameters, and elect the top-performed grey wolves. Then, the algorithm iteratively leverages the enhanced greedy grey wolf social hierarchy as in Algorithm \ref{algo:the_improvement23_gwa} to update the position of each grey wolf, and meanwhile triggers the elimination to eliminate unpromising grey wolves. Upon termination, the best grey wolf represents the optimised IoT resource allocation strategy $\mathcal{A}^O$ of the \emph{BRAD} model. 
The algorithm will firstly utilise the proposed bimetric-balanced and density-aware grey wolf initialiser to initialise the grey wolf population. Meanwhile the algorithm will initialise the parameters, and elect the top-performed grey wolves. Then, the algorithm iteratively leverages the enhanced greedy grey wolf social hierarchy as in Algorithm \ref{algo:the_improvement23_gwa} to update the position of each grey wolf, and meanwhile triggers the elimination to eliminate unpromising grey wolves. Upon termination, the best grey wolf represents the optimised IoT resource allocation strategy $\mathcal{A}^O$ of the \emph{BRAD} model. 

\subsubsection{Optimiser Integration}
\label{sec:section_integration_of_the_grey_wolf_optimiser}

Each grey wolf vector $G_i$ is a $1 \times N$ vector, where $N$ is the number of services to be allocated with IoT resources. Each element in the grey wolf vector $M_i$ has the range of $[0, ub]$, where $ub$ is a constant. If the value lies in $[0, \frac{ub}{2})$, the corresponding service will be allocated to the resource-rich scenario, otherwise, it will be allocated to the resource-scarce scenario. The service execution order is told by the order of values. For each service, if no resource conflict is caused and all its resource requirements can be satisfied, i.e., the equation in Definition $2$ holds, then the service will be allocated with its required resources and be satisfied. 

%%%%%%%%%%%%%%%%%%%%%%%%%%%%%%%%%
%%% Implementation Discussion %%%
%%%%%%%%%%%%%%%%%%%%%%%%%%%%%%%%%

\section{Implementation Discussion}
\label{sec:section_implementation_discussion}

In this section, we introduce how we integrate the concept of Digital Object (DO) abstraction and architecture in our implementation, which includes how it manages and collaborates the IoT resources, as well as its advantages. 

Due to the diversity and heterogeneity of IoT resources \cite{diversity_iot_lan2019iot}, managing them efficiently becomes crucial for effective resource allocation in the smart space. We apply a concept called digital-object-based \cite{doa_sharp2016overview} resource abstraction. The DO-based IoT resource abstraction constitutes an API module, which provides useful APIs to let the IoT resource better interact with external services, such as \textit{show\_availability}, \textit{show\_utilisation\_cost}, etc., and a collaboration module, which is in charge of communicating and collaborating with other resources. For instance, when a service comes and a resource copy is unable to satisfy the request posed by this service, then this resource copy will communicate with other resources of same kind to coordinate the allocation. 

Upon receiving a service resource request, the availability of the DO-based IoT resource will firstly be checked through the API module to see whether it can satisfy this service request. If the request can be satisfied, the API module will communicate its availability to the resource allocation algorithm. If the availability conflicts, it will communicate with the copies of the same resource through the Collaboration module to handle the service request to them. Eventually, the service request will either be satisfied by a resource copy, or unsatisfied due to resource shortage. The API module then reports the allocation outcome to the resource allocator. 

\begin{table*}[!ht]
  \centering
  \begin{tabular}{c|ccccccc}
  \hline
  \multirow{2}{*}{\backslashbox{Value}{Method}} & \multicolumn{2}{c}{Order-based} & \multicolumn{2}{c}{Greedy-based} & \multicolumn{3}{c}{Metaheuristic Algorithm} \\ \cline{2-8} 
                    & RAN & HIT-IHC & GRE\_P & GRE\_O & WOA & FFA & MFO \\ \hline
  Objective & -1038.6 & -729.0 & -1100.7 & -1725.3 & -1761.1 & -1924.8 & -2129.3 \\
  Profit & 525.2 & 459.9 & 542.9 & 586.0 & 879.0 & 1116.4 & 1134.2 \\
  Cost & 1587.2 & 1570.6 & 1613.9 & 1204.6 & 2634.0 & 3639.3 & 3541.6 \\
  Profit/Cost Ratio & 0.33 & 0.30 & 0.34 & 0.50 & 0.34 & 0.31 & 0.32 \\ \hline
  \multirow{2}{*}{\backslashbox{Value}{Method}} & \multicolumn{7}{c}{Metaheuristic Algorithm (continued)}                                               \\ \cline{2-8} 
                    & SCA & PSO & BAT & CS & DE & GWO & GWA \\ \hline
  Objective & -2262.1 & -2323.8 & -2374.2 & -2605.8 & -2661.0 & -2328.0 & \textbf{-3017.2} \\
  Profit & 1218.0 & 1169.4 & 1232.4 & 1249.9 & 1244.9 & 1148.9 & \textbf{1290.14} \\
  Cost & 3828.1 & 3523.1 & 3738.9 & 3643.8 & 3563.6 & 3416.6 & 3433.5 \\
  Profit/Cost Ratio & 0.32 & 0.33 & 0.32 & 0.34 & 0.35 & 0.34 & \textbf{0.38} \\ \hline
  \end{tabular}
  \vspace{1mm}
  \caption{The overall objective, profit, cost and profit/cost ratio of different algorithms. Metrics in which the \emph{BRAD-GWA} algorithm achieves the top-$2$ places are highlighted. }
  \label{tab:objective_profit_cost_pcratio}
  \vspace{-6mm}
\end{table*}

In the \emph{BRAD} model, each IoT resource will be abstracted to form a digital object, we choose to implicitly form each IoT resource as a programming object in the central allocator program, rather than explicitly programming each IoT device. However, the concept works in both ways. Each digital object will interact with the resource allocator through their API module, or collaborate with other digital objects via the collaboration module, forming the digital object architecture (DOA) \cite{doa_sharp2016overview}. By utilising this digital-object-based resource abstraction and the architecture as a whole, it simplifies the interaction with the IoT resources that have heterogeneous characteristics. No matter what type of IoT device is, the defined APIs mask these heterogeneities and abstract all IoT resources to let them be interacted in a unified way. Besides, it promotes easier collaboration and communication between IoT resources during the resource allocation process.

%%%%%%%%%%%%%%%%%%
%%% Experiment %%%
%%%%%%%%%%%%%%%%%%

\section{Performance Evaluation}
\label{sec:section_experiment}

\subsection{Dataset, Parameter and Experimental Setup}
\label{sec:section_dataset_parameter_experimental_setup}

\textbf{Dataset and its Interpretation} We utilise the city hospital and community care home as resource-rich and resource-scarce scenarios of the HIT-IHC system. Each scenario has $10$ kinds of IoT resources that possess heterogeneities. Different IoT resources have different number of copies ranging from $3$ to $50$, which is a typical amount in common IoT applications \cite{meidan2017detection_number_iot_devices}. Besides, each IoT resource copy has different available period and different cost per unit utilisation. Different scenarios also have different IoT resource availabilities, and hence incur diverse utilisation costs. The diverse resource and scenario setting of the HIT-IHC system can testify the robustness of algorithms under the multi-scenario setting. 

In terms of services, we generate $200$ services using Simpy \cite{simpy_documentation}, a discrete event simulator suitable to simulate sequence of IoT resource requests \cite{karanjkar2019simpy}, which also avoids the privacy issues associated with real user service data. The total workloads of these services exceed the capacity of both scenarios, making it persuasive to verify which algorithm can make the wisest choice to allocate services so that the objective is optimised. All services specify its resource requirements with time constraints, and its profit when satisfied in each scenario. 

\textbf{Parameter Setting} The $ub$ parameter of the \emph{BRAD} model is set to be $10$, i.e., all services with their optimised allocation $\mathcal{A}^O_{S^i_n}$ falls in $[0, 5)$ will be allocated to the resource-rich scenario, or otherwise to the resource-scarce scenario. The order of $\mathcal{A}^O_{S^i_n}$ will be used to decide the execution order of services as mentioned in Section \ref{sec:section_integration_of_the_grey_wolf_optimiser}. The profit-cost balancing factor $w$ is set to be $-5$. The allocation upper bound parameter $ub$ of the \emph{BRAD-GWA} algorithm is set to be $10$ which corresponds to the $ub$ parameter in the \emph{BRAD} optimisation model. The default number of search agent is set to $20$. The number of iterations is set to be $100$ to avoid severe computation overhead. 

\textbf{Experimental Setup and Hardware Configuration} The comparing methods we use fall into three categories. Firstly, order-based methods including random allocation (RAN), which allocates IoT resources to services in random order. Another is the default allocation method utilised by the HIT-IHC system as mentioned in Section \ref{sec:section_hit_ihc_intelligent_healthcare_system}, which works in a first come first serve manner, while putting priorities on retired HIT employees. The greedy-based methods including the profit-oriented greedy algorithm (GRE\_P) that sorts services by their highest gainable profit in descending order, then greedily allocate services to be satisfied in the scenario with higher satisfaction profit. Similarly, the objective-oriented greedy strategy (GRE\_O) is used, which prioritises services with highest profit and satisfaction cost ratio. 

The third category is metaheuristic algorithms, which includes Moth-flame Optimiser (MFO) \cite{moth_flame_optimisation_mirjalili2015moth}, Particle Swarm Optimiser (PSO) \cite{kennedy1995particle}, Firefly Algorithm (FFA) \cite{yang2013firefly}, Whale Optimisation Algorithm (WOA) \cite{mirjalili2016whale}, Bat Algorithm (BAT) \cite{yang2012bat}, Sine Cosine Algorithm (SCA) \cite{mirjalili2016sca}, Differential Evolution (DE) \cite{price2013differential}, Cuckoo Search Algorithm (CS) \cite{yang2010engineering}, and vanilla Grey Wolf Optimiser (GWO) \cite{grey_wolf_optimiser_mirjalili2014grey}. To achieve a fair comparison, the profit-cost oriented initialiser is applied for all metaheuristic algorithms, the same number of iterations and search candidates are also applied. 

During experiments, we repeat all experiments $10$ times and the average results are reported. We implement the method using Python $3.8$ and conduct all experiments on a server equipped with Intel i$9$ $9900$K CPU and $32$GB of memory. 

\begin{table*}[!ht]
  \centering
  \begin{tabular}{c|ccccccc}
  \hline
  \multirow{2}{*}{\backslashbox{Value}{Method}} & \multicolumn{2}{c}{Order-based} & \multicolumn{2}{c}{Greedy-based} & \multicolumn{3}{c}{Metaheuristic Algorithm} \\ \cline{2-8} 
                    & RAN & HIT-IHC & GRE\_P & GRE\_O & WOA & FFA & MFO \\ \hline
  Service Allocation & 41.0 & 36.4 & 35.5 & 43.9 & 64.2 & 81.5 & 80.7 \\
  Resource Utilisation & 35\% & 35\% & 36\% & 28\% & 29\% & 31\% & 31\% \\
  Objective/Allocation Ratio & -0.25 & -0.20 & -0.31 & \textbf{-0.40} & -0.28 & -0.24 & -0.26 \\
  Profit/Allocation Ratio & 12.79 & 12.63 & 15.3 & 13.38 & 13.68 & 13.69 & 14.07 \\ 
  Cost/Allocation Ratio & 39.03 & 43.53 & 45.75 & 27.35 & 40.79 & 44.76 & 44.00 \\
  Objective/Utilisation Ratio & -29.8 & -21.1 & -30.58 & -61.77 & -67.36 & -62.18 & -68.61 \\ \hline
  \multirow{2}{*}{\backslashbox{Value}{Method}} & \multicolumn{7}{c}{Metaheuristic Algorithm (continued)}                                               \\ \cline{2-8} 
                    & SCA & PSO & BAT & CS & DE & GWO & GWA \\ \hline
  Service Allocation & 77.9 & 80.9 & 78.4 & 79.4 & 80.8 & 79.1 & \textbf{81.6} \\
  Resource Utilisation & 33\% & 31\% & 33\% & 32\% & 32\% & 31\% & 30\% \\
  Objective/Allocation Ratio & -0.29 & -0.29 & -0.30 & -0.33 & -0.33 & -0.29 & \textbf{-0.37} \\
  Profit/Allocation Ratio & 15.63 & 14.45 & 15.72 & 15.73 & 15.40 & 14.52 & \textbf{15.83} \\ 
  Cost/Allocation Ratio & 49.30 & 43.69 & 48.40 & 45.99 & 44.25 & 43.39 & 42.20 \\
  Objective/Utilisation Ratio & -67.92 & -75.43 & -72.38 & -80.19 & -83.67 & -75.42 & \textbf{-101.59} \\ \hline
  \end{tabular}
  \vspace{1mm}
  \caption{The performance of service allocation and resource utilisation of different algorithms. Metrics in which the \emph{BRAD-GWA} algorithm achieves the top-$2$ places are highlighted. }
  \vspace{-6mm}
  \label{tab:profit_allocation_utilisation}
\end{table*}

\subsection{Performance on Objective Optimisation, Profit and Cost}
\label{sec:comparison_of_profit_and_cost_between_algorithms}

To verify the effectiveness of the \emph{BRAD-GWA} algorithm on objective, profit and cost optimisation, we compare it with $13$ algorithms and present the results in Table \ref{tab:objective_profit_cost_pcratio}. The \emph{BRAD-GWA} algorithm achieves bimetric-balance properly and has the lowest objective value among all methods. Specifically, the \emph{BRAD-GWA} yields $3.14$ times, $74.9\%$, $13.4\%$ and $29.6\%$ objective reduction compared with the default resource allocation method of HIT-IHC, the best performed greedy algorithm GRE\_O, the best performed metaheuristic algorithm DE and the vanilla GWO, respectively. Hence, the effectiveness of \emph{BRAD-GWA} in terms of objective optimisation is verified. 

When optimising the profit, the \emph{BRAD-GWA} algorithm achieves the highest profit performance as in Table \ref{tab:objective_profit_cost_pcratio}, which outperforms other methods by a large margin, i.e., $1.8$ times, $1.2$ times, $3.2\%$ and $12.3\%$ higher than HIT-IHC, GRE\_O, DE and GWO. Hence, the \emph{BRAD-GWA} algorithm is profit-effective. 

However, the best profit performance does not infer the heaviest cost burden. As in Table \ref{tab:objective_profit_cost_pcratio}, the \emph{BRAD-GWA} algorithm incurs the second lowest cost among all metaheuristic algorithms. Its cost is only $0.08\%$ higher than SCA, which possesses the lowest cost. Although the order-based and greedy-based algorithms have a lower cost, they perform poorly in terms of objective optimisation and profit gain. To verify that the \emph{BRAD-GWA} algorithm allocates services in a bimetric-balanced manner, the profit-cost (pc) ratio is calculated. The higher the pc-ratio is, the more profit a unit of cost spent can bring, and hence indicates a more bimetric-balanced service allocation. From Table \ref{tab:objective_profit_cost_pcratio}, the \emph{BRAD-GWA} algorithm enjoys the second highest pc ratio, however, the top-performed GRE\_O algorithm's poor objective and profit performance make it not comparable. Moreover, the \emph{BRAD-GWA} achieves $26.7\%$, $8.6\%$ and $11.8\%$ higher profit-cost ratio compared with the HIT-IHC, best-performed metaheuristic algorithm DE and vanilla GWO, respectively. The significant pc-ratio boost verifies the bimetric-balancing merit of the \emph{BRAD-GWA} algorithm. 

Based on the above analyses, we conclude that the \emph{BRAD-GWA} algorithm is effective in terms of objective minimisation and profit boost. It can also allocate resource to services in a bimetric-balanced way by achieving an excellent profit-cost ratio among compared algorithms. The superior performance over HIT-IHC and the vanilla GWO algorithm also indicates the effectiveness of the advanced \emph{BRAD-GWA} algorithm. 

\begin{figure}[!ht]
  \begin{center}
    \includegraphics[width=0.45\textwidth]{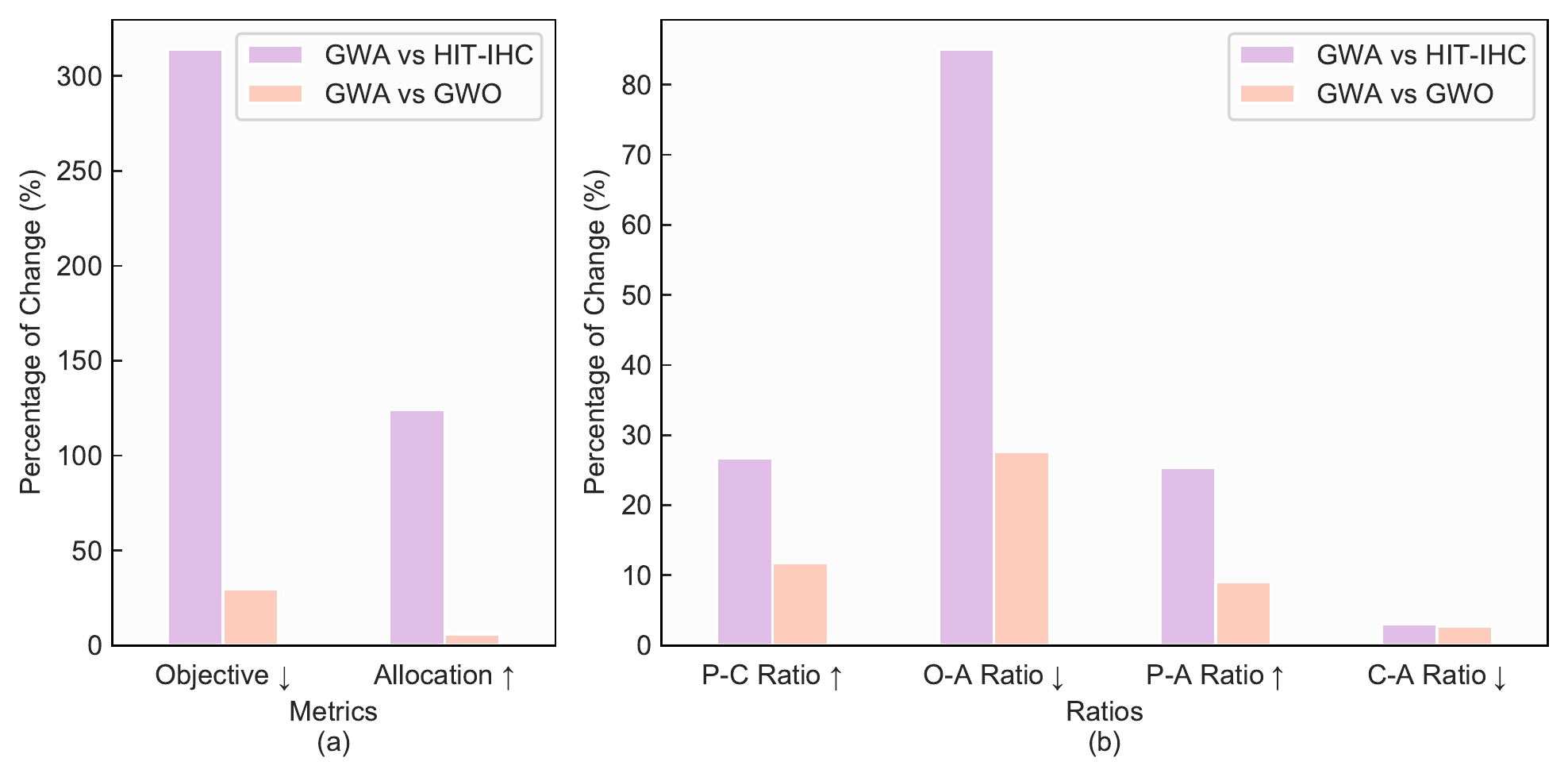}\\
    \caption{(a) The percentage of objective reduction ($\downarrow$) and service allocation boost ($\uparrow$) achieved by the \emph{BRAD-GWA} compared with HIT-IHC and vanilla GWO. (b) The percentage of profit-cost ratio increase, objective-allocation ratio drop, profit-allocation ratio boost, and cost-allocation ratio drop. }
    \vspace{-8mm}
    \label{fig:figure_improvements}
  \end{center}
\end{figure}

\subsection{Performance on Service Allocation and Resource Utilisation}
\label{sec:comparison_of_service_allocation_and_resource_utilisation}

The number of services being satisfied and the average IoT resource utilisation are evaluated and presented in Table \ref{tab:profit_allocation_utilisation}. The \emph{BRAD-GWA} algorithm has the highest number of allocated services, while the order-based and greedy-based algorithms do not perform well on service satisfaction. The objective-allocation ratio reflects the amount of objective each allocated service can bring, the lower the better. The \emph{BRAD-GWA} algorithm achieves the second lowest objective-allocation ratio, second to the GRE\_O, however the latter is not comparable due to poor service satisfaction. Besides, the profit-allocation ratio and the cost-allocation ratio are also computed to indicate the profit and cost each allocated service bring, respectively. The higher the profit-allocation ratio is, the wiser the allocation is, while the lower the cost-allocation ratio is, the better the allocation is, as each service allocation can bring costs as lower as possible. As in Table \ref{tab:profit_allocation_utilisation}, the \emph{BRAD-GWA} yields the highest profit-allocation ratio, outperforming HIT-IHC, best metaheuristic counterpart CS and vanilla GWO by $25.3\%$, $0.6\%$ and $9\%$, respectively. Meanwhile, the \emph{BRAD-GWA} algorithm achieves the second lowest cost-allocation ratio among all metaheuristic methods, while the poor allocation performance makes neither the order-based nor the greedy-based methods comparable. Hence, it verifies that the \emph{BRAD-GWA} algorithm can allocate IoT resources to services in a wise manner and satisfy the most amount of services without causing severe service unsatisfaction and hence possesses the capability to avoid those services that are resource and cost-intensive while being profit-scarce. 

From the resource utilisation aspect, the results of average IoT resource utilisation and the objective-utilisation ratio are presented in Table \ref{tab:profit_allocation_utilisation}. The \emph{BRAD-GWA} yields an average IoT resource utilisation rate of around $30\%$, which is reasonable compared with other methods. In terms of the objective-utilisation ratio, the lower it is, the lower objective value is achieved by one unit of resource utilisation. The \emph{BRAD-GWA} algorithm achieves a significant objective-utilisation ratio drop, which is $3.8$ times, $63.5\%$, $21.4\%$ and $34.7\%$ compared with HIT-IHC, GRE\_O, the best-performed metaheuristic algorithm DE and the vanilla GWO, respectively. Hence, it demonstrates the effectiveness of the \emph{BRAD-GWA} algorithm in terms of utilising resource efficiently to satisfy services. 

\subsection{Comparison with Original HIT-IHC Method and Vanilla GWO Algorithm}
\label{sec:comparison_with_original_hit_ihc_method_and_vanilla_gwo_algorithm}

Finally, we compare the \emph{BRAD-GWA} algorithm with the original method used by HIT-IHC and the vanilla GWO algorithm. As in Figure \ref{fig:figure_improvements}, the \emph{BRAD-GWA} algorithm significantly outperforms the HIT-IHC and vanilla GWO by a large margin in terms of all metrics. Hence, it demonstrates that the proposed \emph{BRAD-GWA} algorithm can significantly improve the effectiveness of the HIT-IHC system, i.e., by allocating IoT resources to services in a wiser bimetric-balanced manner, satisfying more services and utilising resources wisely. Besides, the superior performance achieved by the \emph{BRAD-GWA} over the vanilla GWO also indicates the improvements applied by the \emph{BRAD-GWA} is effective. 

%%%%%%%%%%%%%%%%%%
%%% Conclusion %%%
%%%%%%%%%%%%%%%%%%

\section{Conclusion}
\label{sec:section_conclusion}

In this paper, motivated by problems faced by service resource allocation in IoT-enabled intelligent environments such as the HIT-IHC intelligent healthcare system, we propose the bimetric-balanced resource allocation model \emph{BRAD} which jointly optimises heterogeneous profit and cost model, and considers multi-services with resource-time-constrained requests at the device granularity with deadlock prevention in mind. To effectively manage and collaborate diverse IoT resources, the \emph{BRAD} model utilises the digital-object-based resource abstraction. We utilise the Grey Wolf Optimiser and substantially improve it by fixing several deficiencies and forms the \emph{BRAD-GWA} algorithm. Comprehensive experiments verify the effectiveness of the \emph{BRAD-GWA} algorithm when performing the bimetric-balanced resource allocation for multi-services under the smart IoT space. Specifically, the effectiveness of the HIT-IHC is substantially improved by the \emph{BRAD-GWA} algorithm, and the excellent performance of the \emph{BRAD-GWA} over vanilla GWO also demonstrates the efficacy of the proposed improvements. 

\section*{Acknowledgment}
This work is supported by National Key R\&D Program of China (No.2021YFF0901102). 

\bibliographystyle{IEEEtran}
\bibliography{BRAD}

\end{document}